\documentclass[epj]{svjour}
\usepackage{graphics}
\usepackage{graphicx}% Include figure files
\usepackage{dcolumn}% Align table columns on decimal point
\usepackage{bm}% bold math
\usepackage{CJK}

\begin{document}
\begin{CJK*}{GB}{gbsn}
\title{Scaling of nuclear modification factors for hadrons and light nuclei}
%\author{C. S. Zhou(Öܳ¿Éý)\inst{1,2} \and Y. G. Ma(ÂíÓà¸Õ)\inst{1,a} \and S. Zhang(ÕÅËÉ)\inst{1,b}% etc
\author{C. S. Zhou \inst{1,2} \and Y. G. Ma \inst{1,3} \thanks {Email: ygma@sinap.ac.cn} \and S. Zhang \inst{1}\thanks {Email: zhangsong@sinap.ac.cn}
%is optional - remove next line if not needed
}                     % Do not remove
\offprints{}          % Insert a name or remove this line
\institute{Shanghai Institute of Applied Physics, Chinese Academy of Sciences, Shanghai 201800, China \and University of Chinese Academy of Sciences, Beijing 100049, China
\and ShanghaiTech University, Shanghai 200031, China}
\date{Received: date / Revised version: date}
\abstract{
The number of constituent quarks (NCQ-) scaling of hadrons and the number of constituent nucleons (NCN-) scaling of light nuclei are  proposed  for nuclear modification factors ($R_{cp}$) of hadrons and light nuclei, respectively, according to the experimental investigations in relativistic heavy-ion collisions. Based on coalescence mechanism  the scalings are performed for pions and protons in quark level, and light nuclei $d (\bar d)$ and $^3$He for nucleonic level, respectively, formed in Au + Au and Pb + Pb collisions and nice scaling behaviour emerges.  NCQ or NCN scaling law of $R_{cp}$ can be respectively taken as a probe for quark or nucleon coalescence mechanism for the formation of hadron or light nuclei in relativistic heavy-ion collisions.
\PACS{
      {PACS-key}{25.75.Gz, 12.38.Mh, 24.85.+p}  % \and
      %{PACS-key}{discribing text of that key}
     } % end of PACS codes
} %end of abstract
\maketitle
%
%\modulolinenumbers[5]
%\linenumbers
\section{Introduction}
\label{intro}

Quark-gluon plasma (QGP) which was predicted by the Quantum chromodynamics (QCD) ~\cite{QCD-QGP} can be formed in ultra-relativistic heavy-ion collisions, at the Relativistic Heavy-Ion Collider (RHIC), Brookhaven National Laboratory~\cite{RHICWithePaper-1,RHICWithePaper-2,RHICWithePaper-3,RHICWithePaper-4} and at the Large Hadron Collider (LHC), CERN,  and the  energy dependence of quark matter properties is thought as one of important observables to study various aspects of the QCD phase diagram \cite{Tian,LiuF,NST-1,NST-2,NST-3,NST-4,Ko-1,Ko-2,Ye-1,Ye-2} in the beam energy scan  program at RHIC \cite{Mohanty}.
Extensive experimental results in Au + Au collisions, including particle and anti-particle production and spectra, the nuclear modification factors (NMF) $R_{AA}$ or $R_{cp}$ of hadrons and baryon-to-meson ratios at $\sqrt{s_{NN}}$ = 200 GeV/c have been reported~\cite{RHICWithePaper-1,RHICWithePaper-2,RHICWithePaper-3,RHICWithePaper-4,RHICdata_a,RHICdata_b,Science,Nature,STAR2004}.
Some thermal parameters of the QGP are extracted, for instance the chemical freeze-out temperature $T_{ch}$, the baryon chemical potential $\mu_B$ and the strangeness chemical potential $\mu_S$ etc~\cite{RHICparameter,PBM-nature}, which helps us to understand the phase structure at high temperature.

In the viewpoint of experimental probes, the nuclear modification factor (NMF) $R_{cp}$ is a very useful observable for studying the feature of  QGP~\cite{Rcp-1,Rcp-2}.
Extensive studies have been carried out  both experimentally and theoretically.
These studies indicate that the NMF, which can be represented either by the modification factor between nucleus-nucleus (AA) collisions and proton-proton (pp) collision $R_{AA}$ or the one between the central collisions and peripheral collisions $R_{cp}$, can give   quantitative properties of the nuclear medium response when the high speed jet transverses it. In high transverse momentum ($p_{T}$) region, NMF is suppressed owing to jet quenching effect in hot-dense matter and thus has become one of the robust evidences on the existence of the Quark-Gluon-Plasma~\cite{RHICWithePaper-1,RHICWithePaper-2,RHICWithePaper-3,RHICWithePaper-4,Jet2-1,Jet2-2,Jet2-3,Jet2-4,Jet2-5}. In lower $p_{T}$ region, radial flow boosts or the Cronin effect~\cite{CroninE1-1,CroninE1-2} competes with the quenching effect and enhances the NMF, which has been also demonstrated  by the Relativistic Heavy-Ion Collider (RHIC) beam energy scan (BES) project~\cite{Horvat_BES}. Recently, the study of nuclear modification factor was for the first time extended to intermediate-energy heavy-ion collisions and the radial flow effect on proton NMF has been quantitatively investigated by Ma's group \cite{LvMing-1,LvMing-2}.

It will be of very interesting to check if the $R_{cp}$ of $\pi$ and $p$ can be scaled together by a number of constituent quark (NCQ) scaling law as elliptic flow's NCQ-scaling~\cite{Tian,STAR2004,flow}.
On the other hand, $R_{cp}$ of light nuclei is not yet addressed in relativistic heavy ion collisions.
In the present study, apparent differences between the $R_{cp}$ of light nuclei are found for different nucleon number.
With the help of the coalescence model~\cite{coalescence}, we also found a way to fit them by a number of constituent nucleon (NCN) scaling.
In the following texts, we will discuss these phenomena. 

\section{Number of constituent quark (NCQ) scaling of hadron's $R_{cp}$}

\begin{figure}
\centering
\includegraphics[width=9cm]{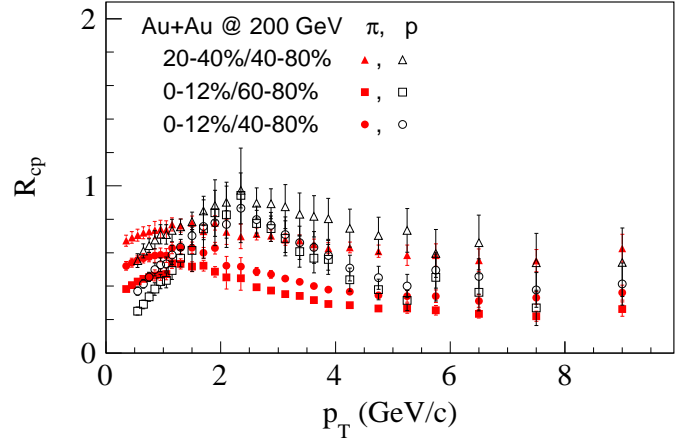}
%\vspace{-0.8cm}
\caption{
(Color online)  $R_{cp}$ for pions and protons in Au+Au collision at $\sqrt{s_{NN}}$ = 200 GeV.  Original data of ${R}_{cp}$  are taken from ref.~ \cite{RHICRcp}.
}
\label{fig:RHICRcp}
\end{figure}

In relativistic heavy ion collisions the production of hadrons at intermediate transverse momentum can be described by coalescence mechanism~\cite{HadronCoal}, i.e.
\begin{eqnarray}
E_h{d^3N_h \over d^3P_h}=B_{n_q} \left(E_q{d^3N_q \over d^3P_q}\right)^{n_q},
\label{eq:coalHadron}
\end{eqnarray}
where $n_q$ is the number of constituent quarks of a hadron and the coefficient $B_{n_q}$ is the probabilities for $n_q$ quarks to hadron coalescence. From the coalescence mechanism it was found that elliptic flow of hadrons can be scaled to the number of constituent quark~\cite{Tian,STAR2004,flow}. And the nuclear modification factor $R_{cp}$ is defined as~\cite{Rcp-1,Rcp-2},
\begin{eqnarray}
R_{cp}(p_T)={[d^2N/p_Tdydp_T/\langle N_{bin}\rangle ]^{central}\over[d^2N/p_Tdydp_T/\langle N_{bin}\rangle ]^{peripheral}},
\label{eq:Rcp}
\end{eqnarray}
where $\langle N_{bin}\rangle$ is the average number of binary nucleon-nucleon collisions per event. The ratio of $\langle N_{bin}\rangle^{central}$ to $\langle N_{bin}\rangle^{peripharal}$ can be found in data~\cite{RHICdata_a} for Au+Au collisions at $\sqrt{s_{NN}}$ = 200 GeV and in data~\cite{ALICE-TAA-1,ALICE-TAA-2} for Pb+Pb collisions at $\sqrt{s_{NN}}$ = 2760 GeV. 
The jets and high $p_{T}$ particles created in the early stage will lose most energy through interactions in the evolution of the system in central A+A collisions and it is confirmed by suppression of $R_{cp}$ at high $p_T$ in experiments in relativistic heavy-ion collisions. Figure~\ref{fig:RHICRcp} displays $R_{cp}$ for pions and protons in three central-peripheral pairs, where the original data are from ref.~\cite{RHICRcp}. Of course, obvious differences between pions and protons are observed.  In this paper, however, we will focus on another topic, i.e. if there is number of constituent quark scaling of $R_{cp}$ for hadrons as that for elliptic flow~\cite{STAR2004,Tian,flow} at intermediate transverse momentum. From eq.~(\ref{eq:coalHadron}) and eq.~(\ref{eq:Rcp}), it can be deduced,
\begin{eqnarray}
R^*_{cp}(p_T) &=&\left(\frac{B_{n_q, central}}{B_{n_q, peripheral}}\right)^{-1/n_q}\left( R_{cp}(n_{q} \cdot p_T)\right)^{1/n_{q}}\nonumber\\
&&\times\left({\langle N_{bin}\rangle ^c \over \langle N_{bin}\rangle ^p}\right)^{1/n_{q}-1},
\label{eq:ScalRcp}
\end{eqnarray}
$R_{cp}^{*}$ is the scaled nuclear modification factor to scale meson's and baryon's $R_{cp}$ together. Unfortunately the coefficient $B_{n_q}$ can not be determined experimentally. So we can try to scale hadron's $R_{cp}$ by the following formula,
\begin{eqnarray}
\widetilde{R}^*_{cp}(p_T) &=&\left( R_{cp}(n_{q} \cdot p_T)\right)^{1/n_{q}}\left({\langle N_{bin}\rangle ^c \over \langle N_{bin}\rangle ^p}\right)^{1/n_{q}-1},\nonumber\\
\widetilde{R}^*_{cp}(p_T)&\equiv&\frac{R^*_{cp}(p_T)}{\left(\frac{B_{n_q, central}}{B_{n_q, peripheral}}\right)^{-1/n_q}}.
\label{eq:ScalRcp_tilde}
\end{eqnarray}

\begin{figure}
\centering
\includegraphics[width=9cm]{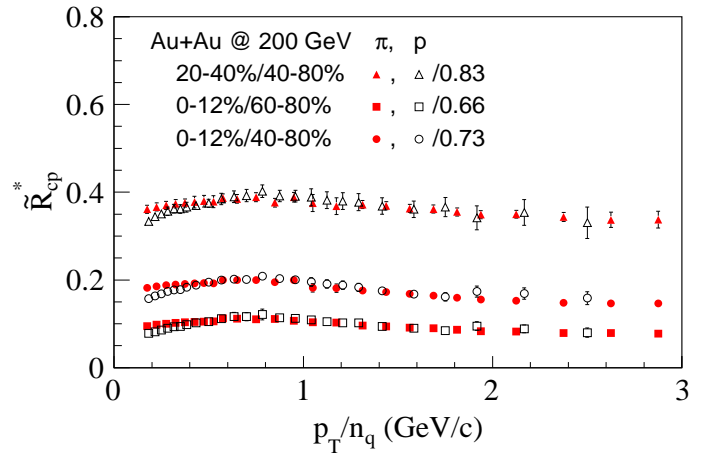}
%\vspace{-0.8cm}
\caption{
(Color online) Number of constituent quark scaling of $R_{cp}$ for pion and proton in Au+Au collision at $\sqrt{s_{NN}}$ = 200 GeV.  Original data of ${R}_{cp}$  are taken from ref.~ \cite{RHICRcp}.
}
\label{fig:RHICRcpScaling}
\end{figure}

\begin{figure*}
\centering
\includegraphics[width=18cm]{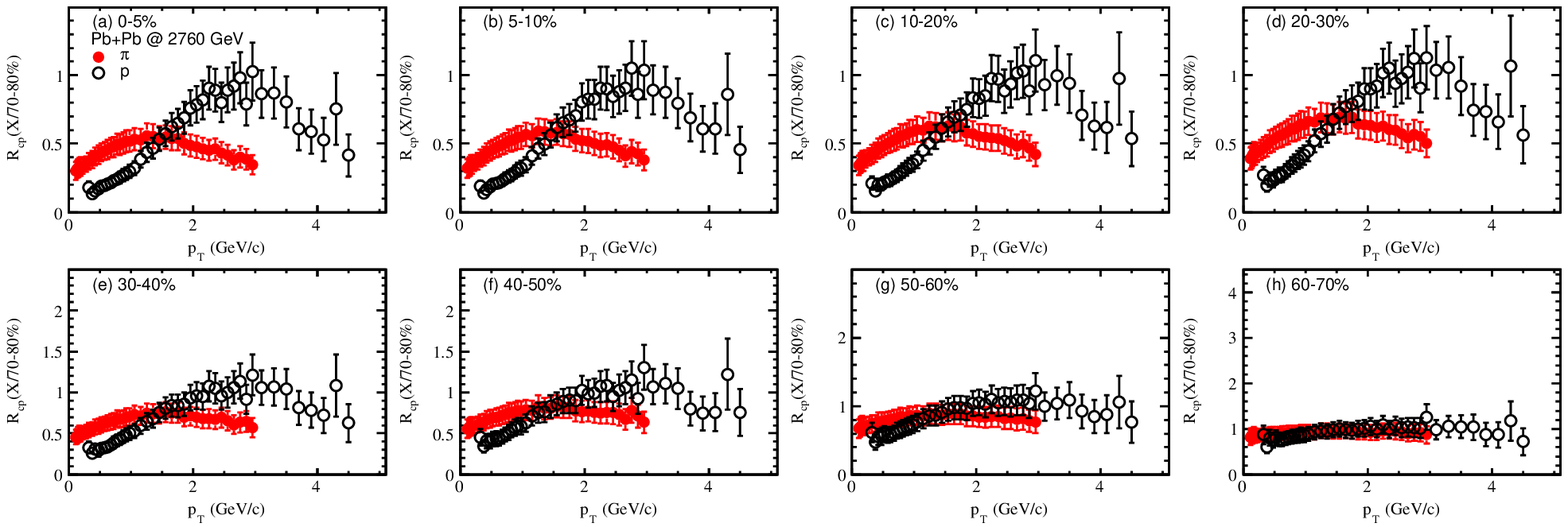}
%\vspace{-0.8cm}
\caption{
(Color online) $R_{cp}$ for pion and proton in Pb+Pb collision at $\sqrt{s_{NN}}$ = 2760 GeV. Original data of ${R}_{cp}$  are taken from refs.~\cite{ALICE-TAA-1,ALICE-TAA-2,ALICERcp-1,ALICERcp-2}.
}
\label{fig:ALICERcp}
\end{figure*}

\begin{figure*}
\centering
\includegraphics[width=18cm]{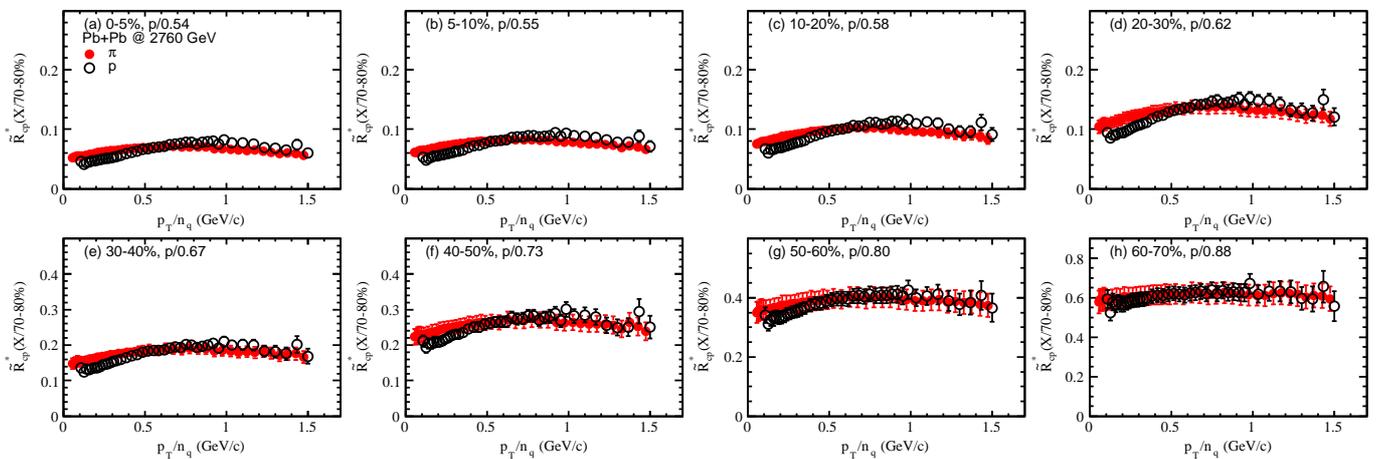}
%\vspace{-0.8cm}
\caption{
(Color online) Number of constituent quark scaling of $R_{cp}$ for pion and proton in Pb+Pb collision at $\sqrt{s_{NN}}$ = 2760 GeV. Original data of ${R}_{cp}$  are taken from refs.~\cite{ALICE-TAA-1,ALICE-TAA-2,ALICERcp-1,ALICERcp-2}.
}
\label{fig:ALICERcpScaling}
\end{figure*}

Figure~\ref{fig:RHICRcpScaling} presents $\widetilde{R}^*_{cp}$ of pion and proton as a function of transverse momentum in Au+Au collisions at $\sqrt{s_{NN}}$ = 200 GeV, where the original data of ${R}_{cp}$  are from ref.~\cite{RHICRcp}. The value of $\widetilde{R}^*_{cp}$ of proton is scaled by a factor listed in the fig.~\ref{fig:RHICRcpScaling}, which can scale pion and proton $\widetilde{R}^*_{cp}$ together. It can be seen from the listed factor which shows the difference between pion's $\widetilde{R}^*_{cp}$ and protons is a constant factor with centrality dependence. 

Similarly, $R_{cp}$ of pions and protons for the Pb+Pb collisions at $\sqrt{s_{NN}}$ = 2760 GeV are plotted as a function of $p_T$ in fig. ~\ref{fig:ALICERcp} for different central-peripheral pair combination. Again, significant differences are observed between pions and protons. After the number of constituent quark scaling for $R_{cp}$, we plot  
fig.~\ref{fig:ALICERcpScaling} for demonstration of $\widetilde{R}^*_{cp}$ of pion and proton, where the original data of ${R}_{cp}$  are from~\cite{ALICE-TAA-1,ALICE-TAA-2,ALICERcp-1,ALICERcp-2}. The scaled factor of $\widetilde{R}^*_{cp}$ for proton is increasing with the increasing  of the centrality of numerator of $\widetilde{R}^*_{cp}$, which takes similar centrality dependence in LHC as in RHIC energy. So it implies that the number of constituent quark scaling of $R_{cp}$ for hadron supports the viewpoint of quark coalescence mechanism for hadron formation in the intermediate transverse momentum range for relativistic heavy-ion collisions and the difference of the scaled $R_{cp}$ between pions and protons keeps a constant factor which displays a centrality dependence.

To summarise this section, we find that the constituent quark scaling holds for the nuclear modification factor of hadrons. It indicates the energy loss of hadrons  is essentially originated from the partonic stage which leads to the same energy loss factor per constituent quark, which is consistent with the concept of quark-gluon plasma. 

\section{Number of constituent nucleon (NCN) scaling for $R_{cp}$ of light nuclei}

\begin{figure*}
\centering
\includegraphics[width=18.0cm]{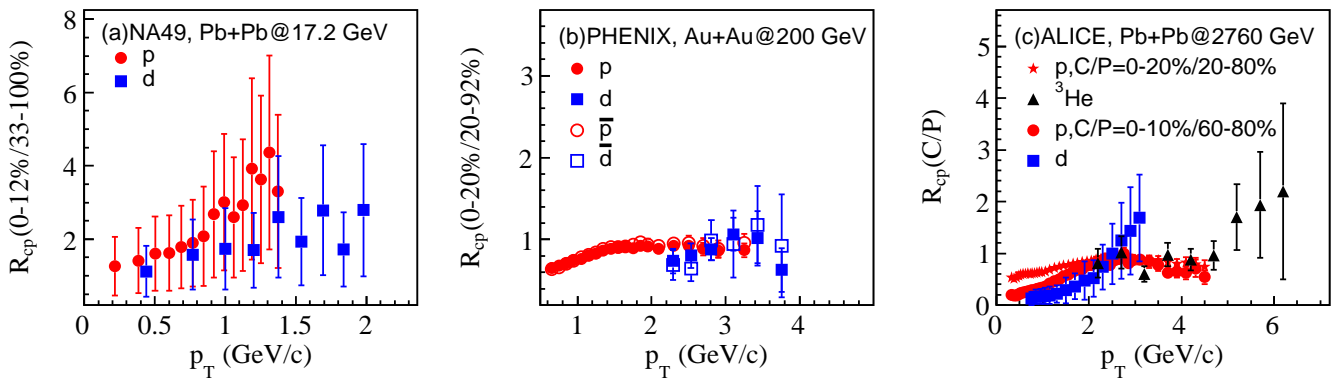}
%\vspace{-0.5cm}
\caption{
(Color online) 
$R_{cp}$ of light nuclei of (a) Pb + Pb at 17.2 GeV, (b) Au + Au at 200 GeV, and (c) Pb + Pb at 2760 GeV, respectively . Original data are taken from the NA49 Collaboration~\cite{LN_NA49} for Pb+Pb collisions at $\sqrt{s_{NN}}$ = 17.2 GeV, the PHENIX Collaboration~\cite{LN_PHENIX-1,LN_PHENIX-2} for Au+Au collisions at $\sqrt{s_{NN}}$ = 200 GeV and the ALICE collaboration~\cite{ALICE-TAA-1,ALICE-TAA-2,ALICERcp-1,ALICERcp-2,LN_ALICE} for  Pb+Pb collisions at $\sqrt{s_{NN}}$ = 2760 GeV.
}
\label{fig:Rcp_nuclei}
\end{figure*}

\begin{figure*}
\centering
\includegraphics[width=18.0cm]{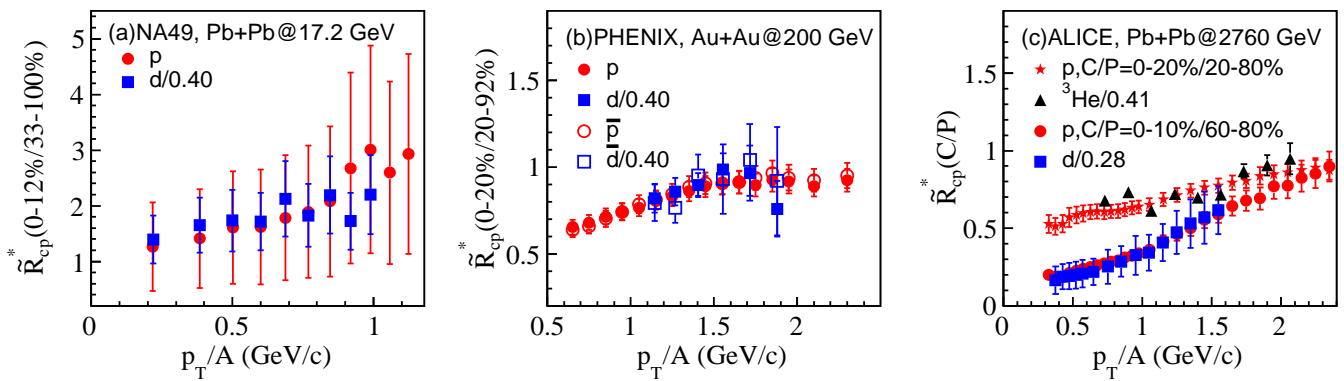}
%\vspace{-0.5cm}
\caption{
(Color online) 
Number of constituent nucleon (NCN) scalings for $R_{cp}$ of light nuclei of (a) Pb + Pb at 17.2 GeV, (b) Au + Au at 200 GeV, and (c) Pb + Pb at 2760 GeV, respectively . Original data are taken from the NA49 Collaboration~\cite{LN_NA49} for Pb+Pb collisions at $\sqrt{s_{NN}}$ = 17.2 GeV, the PHENIX Collaboration~\cite{LN_PHENIX-1,LN_PHENIX-2} for Au+Au collisions at $\sqrt{s_{NN}}$ = 200 GeV and the ALICE collaboration~~\cite{ALICE-TAA-1,ALICE-TAA-2,ALICERcp-1,ALICERcp-2,LN_ALICE} for  Pb+Pb collisions at $\sqrt{s_{NN}}$ = 2760 GeV.
}
\label{fig:Rcp_nuclei_scaling}
\end{figure*}

While the collision system reaches a kinetic freeze-out stage, nucleons could be coalesced to light nuclei. In this case, the coalescence mechanism for the formation of light nuclei is similar to formula~(\ref{eq:coalHadron}) with considering the same distribution of protons and neutrons,
\begin{eqnarray}
E_A{d^3N_A \over d^3P_A}&=&B_A \left(E_p{d^3N_p \over d^3P_p}\right)^Z \left(E_n{d^3N_n \over d^3P_n}\right)^{A-Z},\nonumber\\
&=&B_A \left(E_n{d^3N_n\over d^3P_n}\right)^A,
\label{eq:coalNuclei}
\end{eqnarray}
where
\begin{eqnarray}
P_p=P_n=P_A/A,
\end{eqnarray}
and the coefficient $B_A \sim (1/V)^{A-1}$ is related to the fireball volume in coordinate space and depends on momentum~\cite{coalescence,Xueliang-1,Xueliang-2}. The coefficient $B_A$ can be extracted from data or calculated by coalescence mechanism~\cite{Xueliang-1,Xueliang-2}. The number of constituent nucleon (NCN) scaling of light nuclei's $R_{cp}$ can be obtained through replacing $n_{q}$ by $A$ in eq.~(\ref{eq:ScalRcp}). 

If there exists a similar scaling of $R_{cp}$ for light nuclei as that for elliptic flow~\cite{LowEv2Scaling,Ma_scaling,PRC2016}, 
it certainly supports the coalescence mechanism for formation of light nuclei at kinetic freeze-out stage. Our earlier prediction on number-of-nucleon scaling of elliptic flow \cite{LowEv2Scaling,Ma_scaling} was recently supported by the STAR data from Au + Au collisions \cite{PRC2016}. Even though the energy domain is huge different between our prediction  \cite{LowEv2Scaling,Ma_scaling} and the STAR data  \cite{PRC2016}, the scaling phenomenon is the same, which indicates that the production of light nuclei stems from the nucleonic coalescence mechanism rather than  quark coalescence even in ultra-relativistic energy.  The same concept of  number-of-nucleon scaling of elliptic flow was also followed  by a dynamical model \cite{Oh} as well as a multiphase transport (AMPT) model \cite{Zhu}. 

Here we consider the coalescence mechanism for light nuclei by Eq.~(\ref{eq:coalNuclei}). As the yields of $p(\bar p)$ and $n(\bar n)$ are similar, we have the yields of light nuclei with $A$-mass number obey the Eq.~(\ref{eq:coalNuclei}).
Before the scaling, we show the $R_{cp}$ of light nuclei of  Pb + Pb at 17.2 GeV ~\cite{LN_NA49}, Au + Au at 200 GeV ~\cite{LN_PHENIX-1,LN_PHENIX-2} , and  Pb + Pb at 2760 GeV ~\cite{LN_ALICE,ALICERcp-1,ALICERcp-2,ALICE-TAA-1,ALICE-TAA-2}, respectively. Again, the curves for different light nuclei are not collapsed together. After the scaling for nucleon number, 
fig.~\ref{fig:Rcp_nuclei_scaling} shows the number of constituent nucleon 
on (NCN) scaling for $R_{cp}$ of light nuclei, $\widetilde{R}^*_{cp}$. Figure~\ref{fig:Rcp_nuclei_scaling} (a) represents the $\widetilde{R}^*_{cp}$ of proton and deuteron in Pb + Pb collisions at $\sqrt{s_{NN}}$ = 17.2 GeV and the data are taken from the NA49 collaboration~\cite{LN_NA49}. Figure~\ref{fig:Rcp_nuclei_scaling} (b) gives the $\widetilde{R}^*_{cp}$ of (anti-)proton and (anti)deuteron in Au+Au collisions at $\sqrt{s_{NN}}$ = 200 GeV and the data are from PHENIX collaboration~\cite{LN_PHENIX-1,LN_PHENIX-2}. And fig.~\ref{fig:Rcp_nuclei_scaling} (c) shows the $\widetilde{R}^*_{cp}$ of proton, deuteron and $^{3}$He in Pb + Pb collisions at $\sqrt{s_{NN}}$ = 2760 GeV and the data are from ALICE collaboration~\cite{ALICE-TAA-1,ALICE-TAA-2,ALICERcp-1,ALICERcp-2,LN_ALICE}. From these plots, it was found the nuclear modification factor ($R_{cp}$) of deuteron and $^{3}$He can be scaled to proton's $R_{cp}$ after performing the number of constituent nucleon (NCN) scaling. There is a constant factor for $\widetilde{R}^*_{cp}$ of light nuclei to that of protons. Although the NCN scaling for $R_{cp}$ of light nuclei is studied at three different collision energy $\sqrt{s_{NN}}$ from 17.2 GeV to 2760 GeV, there needs more experimental results of energy and centrality dependence of light nuclei production to explore the constant factor for $\widetilde{R}^*_{cp}$ of light nuclei which is related to the coalescence parameter from eq. (\ref{eq:ScalRcp_tilde}).

In light of this study, we can investigate the $\widetilde{R}^*_{cp}$ for light nuclei experimentally to distinguish their formation mechanism  in heavy ion collisions.  Current data support that the (anti-) light nuclei are coalesced by the (anti-) nucleons. In fact,  the constituent nucleon number scaling of elliptic flow for light nuclei which was firstly proposed in intermediate-energy heavy-ion collisions \cite{LowEv2Scaling,Ma_scaling,Wang-1,Wang-2}  also supports  the coalescence mechanism for the formation of light nuclei at kinetic freeze-out stage in the reaction system. Actually, at RHIC, the elliptic flow of (anti-) light nuclei data of Au + Au at 200 GeV \cite{PRC2016} gives the evidence of the nucleon-number scaling as proposed by Ma's group \cite{LowEv2Scaling}. Furthermore, the observation of anti-hypertriton \cite{Science} and anti-helium 4 \cite{Nature} at STAR also indicates that the anti-nucleon coalescence is responsible for their production yield in the basis of attractive interaction between anti-nucleons \cite{Nature2015-1,Nature2015-2}.

To summarise this section, we find that the constituent quark scaling holds for the nuclear modification factor of light nuclei in relativistic heavy-ion collision. It indicates the energy modification of light nuclei is essentially originated from the hadronic stage rather than protonic stage, which leads to the same energy loss factor per constituent nucleons. On the other hand, in contrast to the scaling behavior of $R_{cp}$ for hadrons, the scaled $R_{cp}$ displays a rising dependence as a function of $p_T$, which could stem from the larger radial flow or multiple nucleon scattering for (anti-) light nuclei as demonstrated in refs.~\cite{LvMing-1,LvMing-2}.

%%%%%%%%%%%%

\section{Conclusion}

We propose, for the first time, that the number of constituent quark (NCQ-) scaling of hadrons and the number of constituent nucleons (NCN-) scaling of  light nuclei for nuclear modification factor $R_{cp}$, respectively, in this work. The NCQ-scaling of $R_{cp}$ of hadrons was applied to the results in Au + Au collisions at $\sqrt{s_{NN}}$ = 200 GeV from the RHIC-STAR and in Pb+Pb collisions at $\sqrt{s_{NN}}$ = 2760 GeV from the LHC-ALICE. The scaled $R_{cp}$ presents a constant difference between pions and protons at intermediate transverse momentum and the constant difference factor shows a centrality dependence trend. This is a supplement for number of constituent quark scaling  besides elliptic flow which supports the coalescence mechanism for formation of hadrons. It is also consistent with the formation of quark-gluon plasma in Au+Au collisions at $\sqrt{s_{NN}}$ = 200 GeV.

Based on the coalescence model, the scaling of $R_{cp}$ has also been tested for light nuclei where the number of constituent nucleon (NCN-) scaling is taken into account. The results display that the $R_{cp}$ spectra of different light nuclei become the one as protons after the scaling. This implies that the scaling factor for $R_{cp}$ of light nuclei should be a constant value and would not be distorted for the structure of $R_{cp}$. This scaling behaviour can be taken as a nucleon coalescence mechanism of light nuclei, and illustrates that those (anti-) light nuclei are  originated from the hadronic stage. 

\vspace{.5cm}
Acknowledgement: This work was supported in part by the Major State Basic Research
Development Program in China under Contract No. 2014CB845400, the
National Natural Science Foundation of China under contract Nos. 11421505, 11220101005, 
11105207 and U1232206, and the CAS Project Grant No. QYZDJSSW-SLH002.

%
% BibTeX users please use
% \bibliographystyle{}
% \bibliography{}

\begin{thebibliography}{}

\bibitem{QCD-QGP} F. Karsch, Nucl. Phys. A \textbf{698}, 199c (2002).

\bibitem{RHICWithePaper-1} I. Arsene \textit{et al.} (BRAHMS Collaboration), Nucl. Phys. A \textbf{757}, 1
(2005).
\bibitem{RHICWithePaper-2} B. B. Back \textit{et al.} (PHOBOS Collaboration), Nucl. Phys. A \textbf{757}, 28 (2005).
\bibitem{RHICWithePaper-3} J. Adames \textit{et al.} (STAR Collaboration), Nucl. Phys. A \textbf{757}, 102 (2005).
\bibitem{RHICWithePaper-4} S. S. Adler \textit{et al.} (PHENIX Collaboration), Nucl. Phys. A \textbf{757}, 184 (2005).


\bibitem{Tian} J. Tian, J. H. Chen, Y. G. Ma, X. Z. Cai, F. Jin, G. L. Ma, S. Zhang and C. Zhong, Phys. Rev. C {\bf 79}, 067901 (2009).

\bibitem{LiuF} N. Yu, F. Liu, K. Wu, Phys. Rev. C {\bf 90}, 024913 (2014).

\bibitem{NST-1} G. Y. Shao, M. Colonna, M. Di Toro, Y. Liu, B. Liu, Nucl. Sci. Techniques {\bf 24}, 050523 (2013).
\bibitem{NST-2} F. M. Liu, Nucl. Sci. Techniques {\bf 24}, 050524 (2013).
\bibitem{NST-3} Y. Hu, Z. Su, W. Zhang, Nucl. Sci. Techniques {\bf 24}, 050522 (2013).
\bibitem{NST-4} H. Wang, Z. Hou, X. Sun,  Nucl. Sci. Techniques {\bf 25}, 040502 (2014).

\bibitem{Ko-1} C. M. Ko, L. W. Chen, V. Greco, F. Li, Z. W. Lin, S. Plumari, T. Song and J. Xu, Nucl. Sci. Techniques {\bf 24}, 050525 (2013).
\bibitem{Ko-2} C. M. Ko and F. Li,  Nucl. Sci. Techniques {\bf 27}, 140 (2016). 

\bibitem{Ye-1}Y. J. Ye, J. H. Chen, Y. G. Ma, S. Zhang, C. Zhong, Phys. Rev. C 93, 044904 (2016).
\bibitem{Ye-2}Y. F. Xu,Y. J. Ye, J. H. Chen,Y. G. Ma, S. Zhang, C. Zhong, Nucl. Sci. Techniques {\bf 27}, 87 (2016).

\bibitem{Mohanty} B. Mohanty (for the STAR Collaboration), J. Phys. G {\bf 38}, 124023 (2011).

\bibitem{RHICdata_a} B.I. Abelev \textit{et al.} (STAR Collaboration), Phys. Lett. B \textbf{655}, 104 (2007).

\bibitem{RHICdata_b} G. Agakishiev \textit{et al.}, Phys. Rev. Lett. \textbf{108}, 072301 (2012).

\bibitem{Science}B. I. Abelev {\it et al.} (STAR Collaboration), Science {\bf 328}, 58 (2010).

\bibitem{Nature}H. Agakishiev {\it et al.} (STAR Collaboration), Nature {\bf 473}, 353 (2011).


\bibitem{STAR2004} J. Adams {\it et al.} (STAR Collaboration), Phys. Rev. Lett. {\bf 92 }, 052302 (2004).


\bibitem{RHICparameter} B.I. Abelev (STAR Collaboration), Phys. Rev. C \textbf{79}, 034909 (2009).

\bibitem{PBM-nature}P. Braun-Munzinger, J. Stachel, Nature {\bf 448}, 302 (2007).


\bibitem{Rcp-1} J. Adams \textit{et al.} (STAR Collaboration), Phys. Rev. Lett. \textbf{91}, 072304 (2003).
\bibitem{Rcp-2} J. Adams \textit{et al.} (STAR Collaboration), Phys. Rev. Lett. \textbf{91}, 172302 (2003).

\bibitem{Jet2-1}J.D. Bjorken, FERMILAB-PUB-82-59-THY and Erratum (unpublished).
\bibitem{Jet2-2}X. N. Wang and M. Gyulassy, Phys. Rev. Lett. {\bf 68}, 1480 (1992).
\bibitem{Jet2-3}E. Wang, X.N. Wang, Phys. Rev. Lett. {\bf 87},  142301 (2001).
\bibitem{Jet2-4}G. L. Ma, Y.G. Ma, S. Zhang {\it et al.}, Phys. Lett. B {\bf 647}, 122 (2007).
\bibitem{Jet2-5}M. Nie and G. Ma, Nucl. Techniques (in Chinese) {\bf 37}, 100519 (2014).


\bibitem{CroninE1-1}J. W. Cronin {\it et al.}, Phys. Rev. Lett. {\bf 31}, 1426 (1973).

\bibitem{CroninE1-2}J. W. Cronin {\it et al.}, Phys. Rev. D {\bf 11}, 3105 (1975).

\bibitem{Horvat_BES} S. P. Horvat {\it et al.} (STAR Collaboration), J. Phys.: Conf. Ser. {\bf 446}, 012017 (2013).

\bibitem{LvMing-1}M. Lv,  Y. G. Ma,  G. Q. Zhang, J. H. Chen, D. Q. Fang, Phys. Lett. B {\bf 733}, 105 (2014).

\bibitem{LvMing-2}M. Lv,  Y. G. Ma,  G. Q. Zhang, J. H. Chen, D. Q. Fang, Nucl. Techniques (in Chinese), {\bf 37}, 100517 (2014).

\bibitem{flow} S. A. Voloshin, Nucl. Phys. A \textbf{715}, 379c (2003).

\bibitem{coalescence} R. Scheibl and U. Heinz, Phys. Rev. C \textbf{59}, 1585 (1999).

\bibitem{HadronCoal} L. Adamczyk {\it et al.} (STAR Collaboration), Phys. Rev. C {\bf  93}, 021903(R) (2016).


\bibitem{ALICE-TAA-1} B. Abelev {\it et al.} (ALICE Collaboration), Phys. Rev. C {\bf 88}, 044909 (2013).

\bibitem{ALICE-TAA-2} B. Abelev {\it et al.} (ALICE Collaboration), Phys. Lett. B {\bf 720}, 52 (2013).


\bibitem{RHICRcp}  B.I. Abelev {\it et al.} (STAR Collaboration), Phys. Rev. Lett. \textbf{97}, 152301 (2006).

\bibitem{ALICERcp-1} B. Abelev {\it et al.} (ALICE Collaboration), Phys. Rev. Lett. \textbf{109}, 252301 (2012).
\bibitem{ALICERcp-2} B. Abelev {\it et al.} (ALICE Collaboration), Phys. Rev. C \textbf{88}, 044910 (2013).


\bibitem{Xueliang-1}L. Xue,  Y. G. Ma, J. H. Chen, S. Zhang, Phys. Rev. C {\bf 85}, 064912 (2012).
\bibitem{Xueliang-2}N.Shah, Y. G. Ma, J. H. Chen, S. Zhang, Phys. Lett. B {\bf 754}, 6 (2016).

\bibitem{LowEv2Scaling} T. Z. Yan, Y. G. Ma, Z. Z. Cai {\it et al.}, Phys. Lett. B {\bf 638}, 50 (2006).

\bibitem{Ma_scaling}Y. G. Ma {\it et al.}, Nucl. Phys. A {\bf 787}, 611c (2007).

\bibitem{PRC2016}L. Adamczyk {\it et al.} (STAR Collaboration), Phys. Rev. C {\bf 94}, 034908 (2016).

\bibitem{Oh} Yongseok Oh and Che Ming Ko,  Phys. Rev. C {\bf 76}, 054910 (2007).

\bibitem{Zhu}Lilin Zhu, Che Ming Ko, Xuejiao Yin, Phys. Rev. C {\bf 92}, 064911 (2015).


\bibitem{LN_NA49} T. Anticic {\it et al.} (NA49 Collaboration), Phys. Rev. C \textbf{69}, 024902 (2004).

\bibitem{LN_PHENIX-1} S. S. Adler {\it et al.} (PHENIX Collaboration), Phys. Rev. C \textbf{69}, 034909 (2004).
\bibitem{LN_PHENIX-2} S. S. Adler {\it et al.} (PHENIX Collaboration), Phys. Rev. Lett. \textbf{94}, 122302 (2005).

\bibitem{LN_ALICE} J. Adam {\it et al.} (ALICE Collaboration), Phys. Rev. C \textbf{93}, 024917 (2016).

\bibitem{Wang-1} J. Wang, Y. G. Ma, G. Q. Zhang, W. Q. Shen, Phys. Rev. C {\bf 90}, 054601 (2014).
\bibitem{Wang-2} J. Wang, Y. G. Ma, G. Q. Zhang, W. Q. Shen, Nucl. Sci. Techniques {\bf 24},   030501 (2013).

\bibitem{Nature2015-1} L. Adamczyk  {\it et al.} (STAR Collaboration), Nature {\bf 527}, 345 (2015).
\bibitem{Nature2015-2} Z. Q. Zhang, Y. G. Ma, Nucl. Sci. Techniques {\bf 27}, 152 (2016). %  doi:10.1007/s41365-016-0147-x

\end{thebibliography}

\end{CJK*}
\end{document}